\documentclass[conference]{IEEEtran}
\IEEEoverridecommandlockouts
\usepackage{cite}
\usepackage{amsmath,amssymb,amsfonts}
\usepackage{bm}
\usepackage{stfloats}
\usepackage[caption=false,font=normalsize,labelfont=sf,textfont=sf]{subfig}

\usepackage{algorithmic}
\usepackage{graphicx}
\usepackage{textcomp}
\usepackage{xcolor}
\def\BibTeX{{\rm B\kern-.05em{\sc i\kern-.025em b}\kern-.08em
    T\kern-.1667em\lower.7ex\hbox{E}\kern-.125emX}}
\begin{document}

\title{Vision Transformer for Adaptive Image Transmission over MIMO Channels \\
}

 \author{%
   \IEEEauthorblockN{Haotian Wu, Yulin Shao,
                    Chenghong Bian,
                     Krystian Mikolajczyk,
                     and Deniz Gündüz}
   
   \IEEEauthorblockA{Department of Electrical and Electronic Engineering, Imperial College London, London SW7 2BT, UK\\
    Email:$\left\{haotian.wu17, y.shao, c.bian22, k.mikolajczyk, d.gunduz\right\}$ @imperial.ac.uk}

}

\maketitle

\begin{abstract}
This paper presents a vision transformer (ViT) based joint source and channel coding (JSCC) scheme for wireless image transmission over multiple-input multiple-output (MIMO) systems, called ViT-MIMO. The proposed ViT-MIMO architecture, in addition to outperforming separation-based benchmarks, can flexibly adapt to different channel conditions without requiring retraining. Specifically, exploiting the self-attention mechanism of the ViT enables the proposed ViT-MIMO model to adaptively learn the feature mapping and power allocation based on the source image and channel conditions. Numerical experiments show that ViT-MIMO can significantly improve the transmission quality across a large variety of scenarios, including varying channel conditions, making it an attractive solution for emerging semantic communication systems.
\end{abstract}


\section{Introduction}
The design of efficient image communication systems over wireless channels has recently attracted a lot of interest due to the increasing number of Internet-of-things (IoT) and edge intelligence applications\cite{gunduz2020communicate,FLOAC}. 
The traditional solution from Shannon's separation theorem is to design source and channel coding independently; however, the separation-based approach is known to be sub-optimal in practice, which becomes particularly limiting in applications that impose strict latency constraints\cite{jankowski2020wireless}. Despite the known theoretical benefits, designing practical joint source channel coding (JSCC) schemes has been an ongoing challenge for many decades. Significant progress has been made in this direction over the last years thanks to the introduction of deep neural networks (DNNs) for the design of JSCC schemes\cite{bourtsoulatze2019deep,kurka2020deepjscc,choi2019neural,tung2022deepwive,DTAT}. The first deep learning based JSCC (DeepJSCC) scheme for wireless image transmission is presented in \cite{bourtsoulatze2019deep}, and it is shwon to outperform the concatenation of state-of-the-art image compression algorithm better portable graphics (BPG) with LDPC codes. It was later extended to transmission with adaptive channel bandwidth in\cite{kurka2021bandwidth} and to the transmission over multipath fading channels in \cite{yang2021ofdm} and \cite{wu2022channel}.

To the best of our knowledge, all the existing papers on DeepJSCC consider single-antenna transmitters and receivers. While there is a growing literature successfully employing DNNs for various multiple-input multiple-output (MIMO) related tasks, such as detection, channel estimation, or channel state feedback \cite{samuel2019learning,8353153,bian2022learning,mashhadi2021pruning,he2020model}, no previous work has so far applied DeepJSCC to the more challenging MIMO channels. MIMO systems are known to boost the throughput and spectral efficiency of wireless communications, showing significant improvements in the capacity and reliability. JSCC over MIMO channels is studied in \cite{gunduz2008joint} from a theoretical perspective. It is challenging to design a practical JSCC scheme for MIMO channels, where the model needs to retrieve coupled signals from different antennas experiencing different channel gains. A limited number of papers focus on DNN-based end-to-end MIMO communication schemes. The first autoencoder-based end-to-end MIMO communication method is introduced in\cite{o2017deep}. In \cite{song2020benchmarking}, the authors set the symbol error rate benchmarks for MIMO channels by evaluating several AE-based models with channel state information (CSI). A singular-value decomposition (SVD) based autoencoder is proposed in \cite{zhang2022svd} to achieve the state-of-the-art bit error rate. However, these MIMO schemes only consider the transmission of bits at a fixed signal-to-noise ratio (SNR) value, ignoring the source signal's semantic context and channel adaptability.

In this paper, we design an end-to-end unified DeepJSCC scheme for MIMO image transmission. In particular, we introduce a vision transformer (ViT) based DeepJSCC scheme for MIMO systems with CSI, called ViT-MIMO. Inspired by the success of the attention mechanism in the design of flexible communication schemes\cite{wu2022channel,shao2022attentioncode,ozfatura2022all,xu2021wireless}, we leverage the self-attention mechanism of the ViT in wireless image transmission. Specifically, we represent the channel conditions with a channel heatmap, and adapt the JSCC encoding and decoding parameters according to this heatmap. Our method can learn global attention between the source image and the channel conditions in all the intermediate layers of the DeepJSCC encoder and decoder. Intuitively, we expect this design to simultaneously learn feature mapping and power allocation based on the source semantics and the channel conditions. 
Our main contributions can be listed as follows:
\begin{itemize}
\setlength\itemsep{-.3em}
\item To the best of the authors’ knowledge, our ViT-MIMO is the first DeepJSCC-enabled MIMO communication system for image transmission, where a ViT is designed to explore the contextual semantic features of the image as well as the CSI in a self-attention fashion.
\vspace{0.1cm}

\item Numerical results show that our ViT-MIMO model significantly improves the transmission quality over a large range of channel conditions and bandwidth ratios, compared with the traditional separate source and channel coding schemes adopting BPG image compression algorithm with capacity achieving channel transmission. In addition, our proposed ViT-MIMO is a flexible end-to-end model that can adapt to varying channel conditions without retraining.
\end{itemize}

\section{System Model}
We consider a $M\times M$ MIMO communication system, where an $M$-antenna transmitter aims to deliver an image $\bm{S}\in \mathbb{R}^{h\times w\times 3}$ to a $M$-antenna receiver ($h$ and $w$ denote the height and width of the image, while $3$ refers to the color channels R, G and B). The transmitter encodes the image into a vector of channel symbols $\bm{X}\in \mathbb{C}^{M\times k}$, where $k$ is the number of channel uses. Following the standard definition \cite{kurka2020deepjscc}, we denote the \textit{bandwidth ratio} (i.e., channel usage to the number of source symbols ratio) by $R\triangleq k/n$, where $n=3hw$ is the number of source symbols. The transmitted signal $\bm{X}$ is subject to a power constraint $P_s$:
\begin{equation}
   \frac{1}{Mk}\|\bm{X}\|_F^2\leq P_s,
   \label{power}
\end{equation}
where $\|\cdot\|_\text{F}$ denotes the Frobenius norm, and we set $P_s=1$ without loss of generality.

The channel model can be written as: 
\begin{equation}
   \bm{{Y}}=\bm{H}\bm{X}+\bm{W},
\label{MIMO_channel}
\end{equation}
where $\bm{X}\in \mathbb{C}^{M\times k}$ and $\bm{Y}\in \mathbb{C}^{M\times k}$ denote the channel input and output matrices, respectively, while $\bm{W}\in \mathbb{C}^{M\times k}$ is additive white Gaussian noise (AWGN) term that follows independent and identically distributed (i.i.d.) complex Gaussian distribution with zero mean and $\sigma_w^2$ variance, i.e., $\bm{W}[i,j] \sim \mathcal{CN}(0,\sigma_w^2)$. The entries of the channel gain matrix $\bm{H}\in \mathbb{C}^{M\times M}$ follow i.i.d. complex Gaussian distribution with zero mean and variance $\sigma_h^2$, i.e., $\bm{H}[i,j]\sim \mathcal{CN}(0,\sigma_h^2)$. We consider a slow block-fading channel model, in which the channel matrix $\bm{H}$ remains constant for $k$ channel symbols, which corresponds to the transmission of one image and takes an independent realization in the next block. We assume that the CSI is known both by the transmitter (CSIT) and the receiver (CSIR). 

Given the channel output $\bm{Y}$, receiver reconstructs the  source image as $\bm{\hat{S}}\in \mathbb{R}^{h\times w\times 3}$. We use the peak signal-to-noise ratio (PSNR) as the distortion metric:
\begin{equation}
\text{PSNR}=10\log_{10}\frac{\|\bm{S}\|_\infty^2}{\mathbb{E}\|\bm{S}-\bm{\hat{S}}\|^2_F}~(\text{dB}),
\label{psnr}
\end{equation}
where $\|\cdot\|_\infty$ is the infinity norm and the expectation is computed over each pixel for mean squared error (MSE).

\begin{figure}[t] 
    \centering
    \includegraphics[scale=0.47]{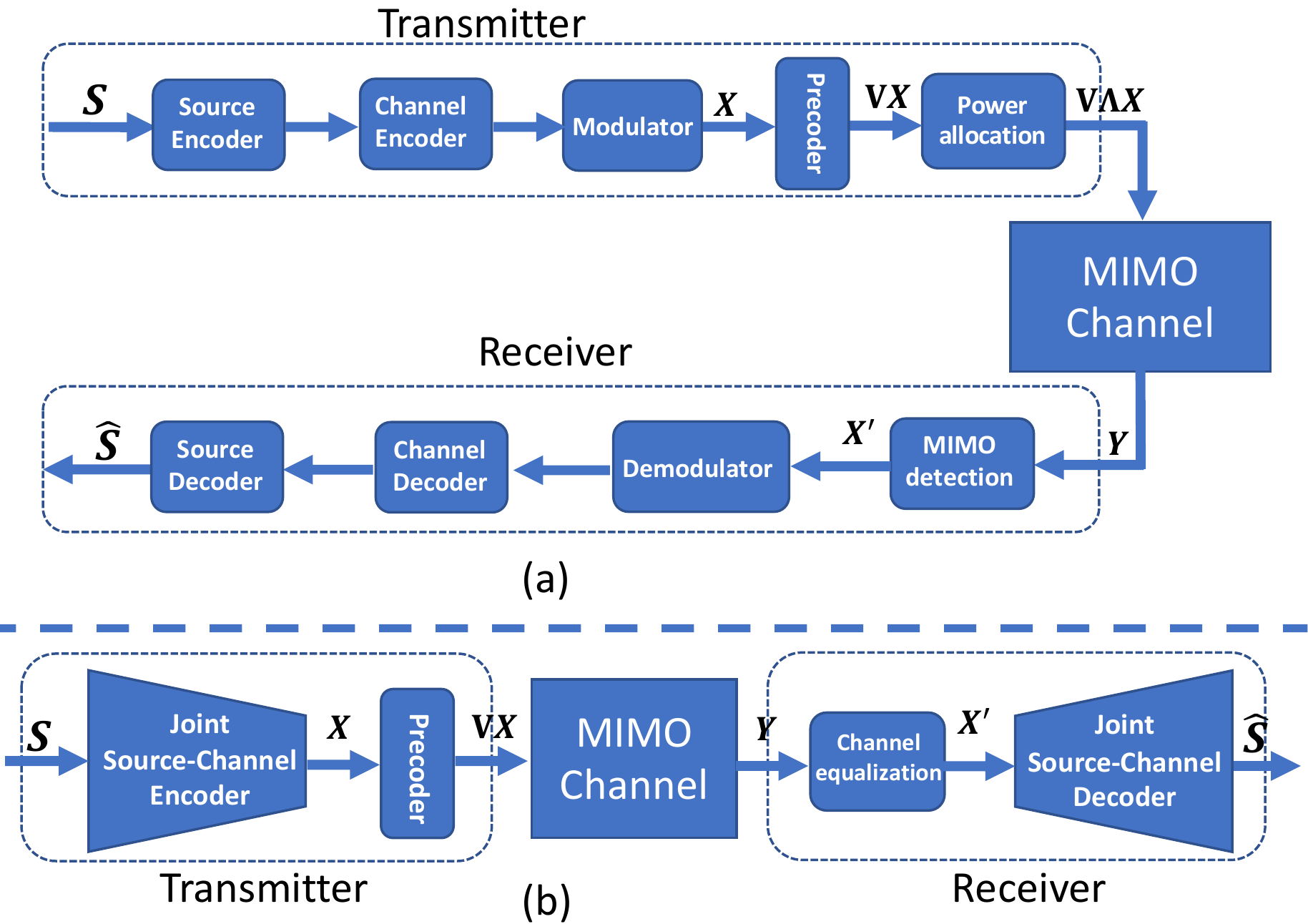}
    \caption{Block diagram of the MIMO image transmission system: (a) conventional separate source-channel coding scheme, and (b) deep learning based JSCC scheme.}
    \label{MIMO_pipeline}
\end{figure}

As shown in Fig. \ref{MIMO_pipeline}, there can be two approaches to solve this problem. The traditional separate source and channel coding scheme and JSCC.

\subsection{Separate source and channel coding}
For the traditional separate scheme, we sequentially perform image compression, channel coding, and modulation to generate the channel input matrix $\bm{X}$, the elements of which are constellations with average power normalized to 1. Specifically, given the CSI, we first decompose the channel matrix by singular-value decomposition (SVD), yielding $\bm{{H}}=\bm{U\Sigma V^H}$, where $\bm{U}\in \mathbb{C}^{M\times M}$ and $\bm{V}\in \mathbb{C}^{M\times M}$ are unitary matrices, and $\bm{\Sigma}$ is a diagonal matrix whose singular values are in descending order.
We denote $\bm{\Sigma}$ by $\text{diag}(s_1,s_2,\dots, s_M)$, where $s_1\geq \dots \geq s_M$.

Let us denote the power allocation matrix by $\bm{\Lambda}$, where $\bm{\Lambda}$ is diagonal with its diagonal element being the power allocation weight for the signal stream of each antenna. $\bm{\Lambda}$ can be derived from standard water-filling algorithms. With power allocation and SVD precoding (precoding $\bm{X}$ into $\bm{VX}$), \eqref{MIMO_channel} can be rewritten as
\begin{equation}\label{eq:MIMO1}
   \bm{{Y}}=\bm{H}\bm{V}\bm{\Lambda X}+\bm{W}=\bm{U\Sigma}\bm{\Lambda X}+\bm{W}. 
\end{equation}
Multiplying both sides of  \eqref{eq:MIMO1} by $\bm{U}^H$ (MIMO detection) gives
\begin{equation}
   \bm{X'}=\bm{\Sigma}\bm{\Lambda X}+\bm{U}^{H}\bm{W}.
   \label{svd_MIMO_out}
\end{equation}

As can be seen, SVD-based precoding converts the MIMO channel into a set of parallel subchannels with different SNRs.
In particular, SNR of the $i$-th subchannel is determined by the $i$-th singular value of $s_i$ and the $i$-th power allocation coefficient of $\bm{\Lambda}$. 

Given $\bm{X'}\in \mathbb{C}^{M\times k}$, the receiver performs demodulation, channel decoding, and source decoding sequentially to reconstruct the source image as $\bm{\hat{S}}$.

\subsection{DeepJSCC}
In DeepJSCC, we exploit deep learning to parameterize the encoder and decoder functions, which are trained jointly on an image dataset and the channel model in \eqref{MIMO_channel}. 
Let us denote the DeepJSCC encoder and decoder by $f_{\bm{\theta}}$ and $f_{\bm{\phi}}$, respectively, where $\bm{\theta}$ and $\bm{\phi}$ denote the network parameters. We have
\begin{equation}
  \bm{X}=f_{\bm{\theta}}(\bm{S},\bm{H},\sigma_w^2).
   \label{dl_encoder}
\end{equation}
Unlike the separate source-channel coding scheme, the transmitter does away with power allocation. Instead, we leverage the DeepJSCC encoder to perform feature mapping and power allocation. Intuitively, the DNN is expected to transmit critical features over subchannels with higher SNRs, thereby improving the image reconstruction performance.

We note here that one option is to train the encoder/decoder networks directly, hoping they will learn to exploit the spatial degrees-of-freedom MIMO channel provides. We will instead follow the model-driven approach, where we will exploit the SVD as is done above for the channel coding scheme, and convert the MIMO channel into subchannels, which will then be used for training the JSCC encoder/decoder pair. The received signal can be written as
\begin{equation}
\bm{{Y}}=\bm{HVX+W}=\bm{U\Sigma}\bm{X}+\bm{W}.
\label{precode_dl_MIMO}
\end{equation}

To simplify the network training, we apply MIMO equalization by left multiplying both sides of  \eqref{precode_dl_MIMO} by $\bm{\Sigma}^{\dagger}\bm{U}^H$ to obtain
\begin{equation}
\bm{X'}=\bm{\Sigma}^{\dagger}\bm{U}^{H}\bm{Y}=\bm{X}+\bm{W'},
\label{MIMO_detec}
\end{equation}
where $\bm{W}'\triangleq \bm{\Sigma^{\dagger}U^HW}\in \mathbb{C}^{M\times k}$ is the equivalent noise term and $\bm{\Sigma^{\dagger}}$ is the Moore-Penrose inverse of matrix $\bm{\Sigma}$.

Finally, we feed both $\bm{X'}$ and the CSI into the DeepJSCC decoder to recover the image as
\begin{equation}
 \bm{\hat{S}}=f_{\bm{\phi}}(\bm{X'},\bm{H},\sigma_w^2).
   \label{dl_decoder}
\end{equation}

\section{Methodology}
\begin{figure}[t] 
    \centering
    \includegraphics[scale=0.58]{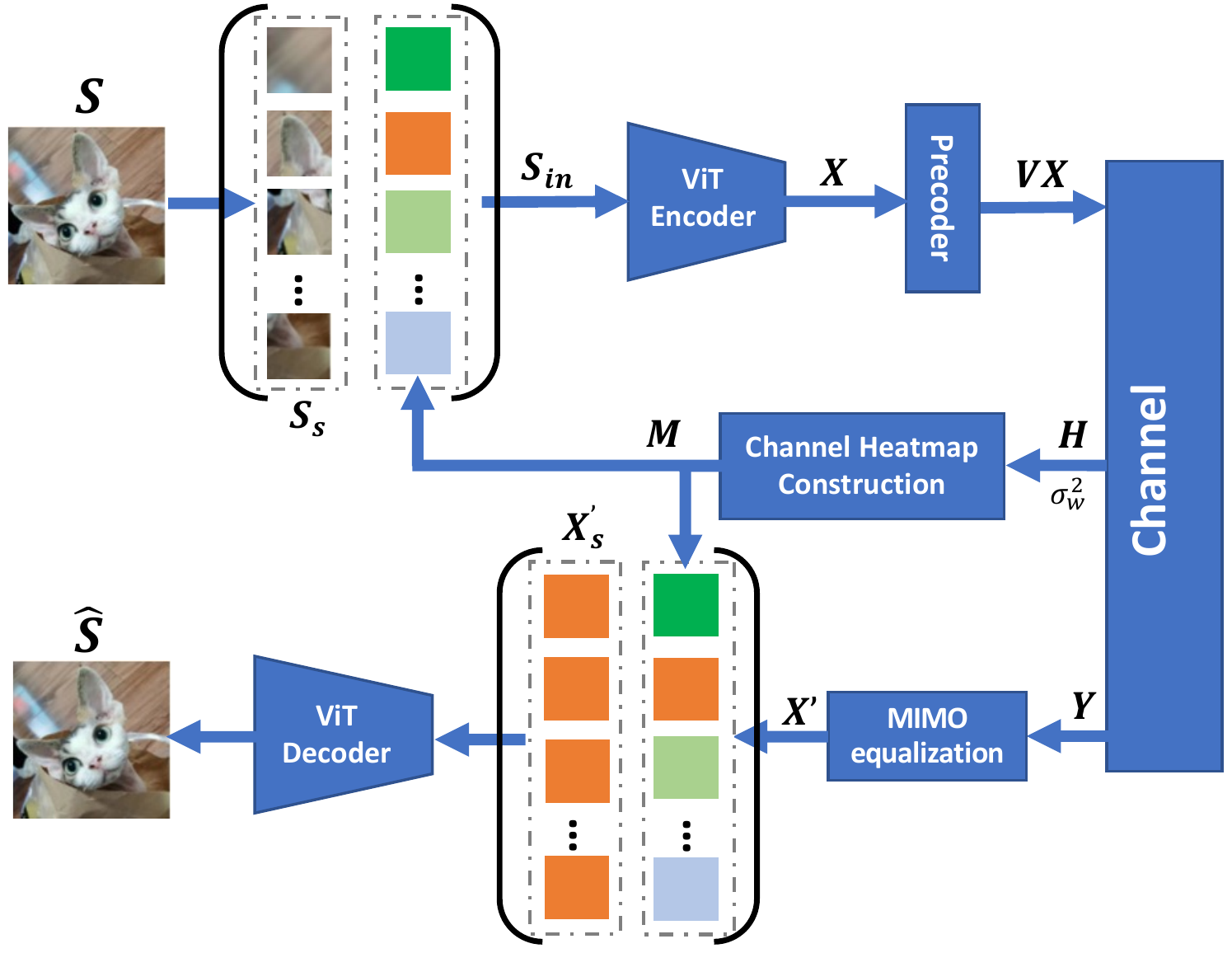}
    \caption{The pipeline of our ViT-MIMO scheme.}
    \label{Secom_pipeline}
\end{figure}
\begin{figure*}[tb] 
    \centering
    \includegraphics[scale=0.425]{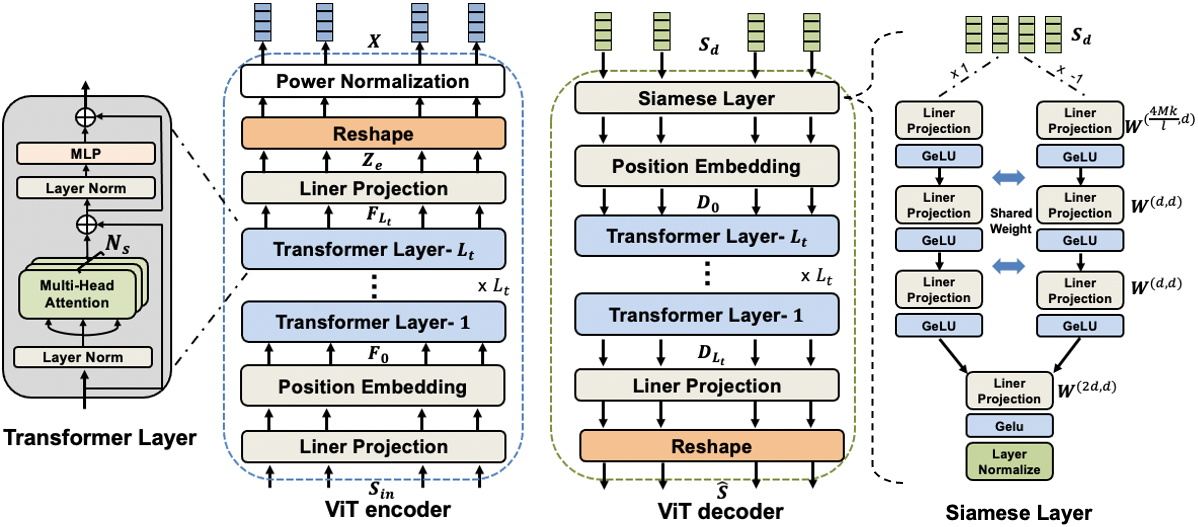}
    \caption{Inner structures of the ViT encoder and decoder architectures.}
    \label{ViT_structure}
\end{figure*}

To enable efficient image transmission in MIMO systems, we propose a new DeepJSCC architecture, dubbed ViT-MIMO, in this section. The pipeline of our ViT-MIMO scheme is illustrated in Fig. \ref{Secom_pipeline}.
In the big picture, ViT-MIMO utilizes a pair of ViTs as the encoder $f_{\bm{\theta}}$ and decoder $f_{\bm{\phi}}$, respectively, which have symmetric inner structures, as detailed later in Fig. \ref{ViT_structure}. 
In particular, the CSI ($\bm{H},\sigma_w^2$) is fed into the encoder and decoder in the form of a ``heatmap''.
In the following, we detail the pipeline of ViT-MIMO in five main steps: image-to-sequence transformation, channel heatmap construction, ViT encoding, ViT decoding, and the loss function.

\subsection{Image-to-sequence transformation}
To construct the input of ViT-MIMO, we first convert the three-dimensional input image $\bm{S}$ into a sequence of vectors, denoted by $\bm{S_s}$. Specifically, given a source image $\bm{S}\in\mathbb{R}^{h\times w\times 3}$, we divide $\bm{S}$ into a grid of $p\times p$ patches, and flatten the pixel intensities of each patch to form a sequence of vectors of dimension $\mathbb{R}^{\frac{3hw}{p^2}}$. In this way, $\bm{S}$ is converted to $\bm{S_s}\in\mathbb{R}^{l\times c}$, where $l={p^2}$ is the sequence length and $c\triangleq\frac{3hw}{p^2}$ is the dimension of each vector.
\subsection{Channel heatmap construction}
To enable efficient training, we construct a channel heatmap from CSI, indicating the effective noise variance faced by each channel symbol generated by ViT.


Let us denote the average power of $\bm{W'}$ as:
\begin{equation}
\bm{P_n} \triangleq \bm{\sigma_w^2\Sigma^{\dagger}U^H}\bm{J},
\end{equation}
where $\bm{J}\in \mathbb{R}^{M\times k}$ is the matrix of ones. 

The heatmap $\bm{M}\in\mathcal{R}^{l\times \frac{2Mk}{l}}$ is constructed as follows:
\begin{equation}
   \bm{{M}}\triangleq \textit{reshape} \left(\textit{concat}\left(\frac{1}{2}\bm{P_n},\frac{1}{2}\bm{P_n}\right)\right),
   \label{heatmap_open}
\end{equation}
where $\textit{concat}(\cdot)$ and $\textit{reshape}(\cdot)$ denote the concatenation and reshape operations, respectively. We concatenate two matrices $\frac{1}{2}\bm{P_n}$ to get the same shape and equivalent noise term faced by the real encoder output.

We expect that feeding the CSI with this heatmap will simplify the training process as the model only needs to focus on the `additive' noise power faced by channel symbols. 

\subsection{ViT-encoding}
The input to our encoder $f_{\bm{\theta}}$, $\bm{{S}_{in}}$, is the concatenation of $\bm{S_s}$ and $\bm{M}$:
\begin{equation}
   \bm{{S}_{in}}=\textit{concat}(\bm{S_s},\bm{M})\in \mathbb{R}^{l\times (c+\frac{2Mk}{l})}.
\label{ViT_input}
\end{equation}

The architecture of our encoder is shown in Fig. \ref{ViT_structure}, mainly consisting of linear projection layers, a position embedding layer, and several transformer layers. 

\textbf{Linear projection and position embedding:}
$\bm{S_{in}}$ firstly goes through a linear projection operation with parameters $\bm{W_{0}}\in \mathbb{R}^{(c+\frac{2Mk}{l})\times d}$ and a position embedding operation $P_{e}(\cdot)$ to get the initial input $\bm{F_{0}}\in \mathbb{R}^{l\times d}$ for the following transformer layers as:
\begin{equation}
\bm{F_{0}}=\bm{S_{in}}\bm{W_{0}}+P_{e}(\bm{p}),
\label{pos_embed}
\end{equation}
where $d$ is the output dimension of the hidden projection layer, and $P_{e}(\cdot)$ is a dense layer that embeds the index vector $\bm{p}$ of each patch into a $d$ dimensional vector.

\textbf{Transformer Layer:}
As shown in Fig. \ref{ViT_structure}, the intermediate feature map $\bm{F_{i}}$ is generated by the $i\textit{-}th$ transformer layer by a multi-head self-attention (MSA) block and a multi-layer perceptron (MLP) layer as:
\begin{equation}
\bm{F_{i}}=MSA(\bm{F_{i-1}})+MLP(MSA(\bm{F_{i-1}})),
\label{trans_layer_eqn}
\end{equation}
where $\bm{F_{i}}\in \mathbb{R}^{l\times d}$ is the output sequence of the $i$-th transformer layer, GeLU activation and layer normalization operations are applied before each MSA and MLP block. 

Each MSA block consists of $N_{s}$ self-attention (SA) modules with a residual skip, which can be formulated as:
\begin{equation}
MSA(\bm{F_{i}})=\bm{F_{i}}+[SA_1(\bm{F_{i}}),\cdots,SA_{N_{s}}(\bm{F_{i}})]\bm{W_i},
\label{msa}
\end{equation}
where the output of all SA modules $SA(\bm{F_{i}})\in \mathbb{R}^{l\times d_{s}}$ are concatenated for a linear projection $\bm{W_i}\in \mathbb{R}^{d_{s}N_{s}\times d}$, $d_s=d/N_{s}$ is the output dimension of each SA operation.

For each SA module, the operations can be formulated as \eqref{SA_eqn}:
\begin{equation}
SA(\bm{F_{l-1}})=softmax(\frac{\bm{qk}^{T}}{\sqrt{d}})\bm{v},
\label{SA_eqn}
\end{equation}
where $\bm{q},\bm{k},\bm{v}\in \mathbb{R}^{l\times d_s}$ are the query, key, and value vectors, respectively,  generated from three linear projection layers $\bm{W_q,W_k,W_v}\in \mathbb{R}^{d\times d_s}$ as:
\begin{equation}
    \bm{q}=\bm{F_{l-1}W_q},~~\bm{k}=\bm{F_{l-1}W_k},~~\bm{v}=\bm{F_{l-1}W_v}.
    \label{sa}
\end{equation}

\textbf{Linear projection and power normalization:}
After $L_t$ transformer layers, we apply a linear projection $\bm{W_c}\in \mathbb{R}^{d\times \frac{2Mk}{l}}$ to map the output of the transformer layers $\bm{F_{L_t}}$ into the channel symbols as: 
\begin{equation}
\bm{Z_e}=\bm{F_{L_t}}\bm{W_c},
\label{last_dense}
\end{equation}
where $\bm{Z_e}\in \mathbb{R}^{l\times \frac{2Mk}{l}}$ is then reshaped and normalized to satisfy the power constraints to form the complex channel input symbols $\bm{X}\in \mathbb{C}^{M\times k}$.

\subsection{ViT decoding}
Before decoding the received signals, we perform an equalization operation as in \eqref{MIMO_detec} on the channel output $\bm{Y}$ to get $\bm{X'}\in \mathbb{C}^{M\times k}$, which is then reshaped into $\bm{X^{'}_{s}} \in \mathbb{R}^{l\times \frac{2Mk}{l}}$ as the input of the decoder. With the equalized signal $\bm{X^{'}_{s}}$ and the noise heatmap $\bm{M}$, a ViT-based decoder $f_{\bm{\phi}}$ is designed to recover the source image $\bm{\hat{S}}=f_{\bm{\phi}}(\bm{X^{'}_{s}},\bm{M})$, which consists of a Siamese layer, position embedding layer, transformer layer and linear projection layer.

\textbf{Siamese layer and position embedding:} We design a weight-shared Siamese layer $\text{Siam}(\cdot)$ consisting of several linear projection layers and GeLU activation functions. To form the input of $\text{Siam}(\cdot)$, we concatenate $\bm{X^{'}_{s}}$ and $\bm{M}$ as: 
\begin{equation}
 \bm{S_{d}}=\textit{concate}(\bm{X^{'}_{s}},\bm{M})\in \mathbb{R}^{l\times \frac{4Mk}{l}}.  
\end{equation}
Then $\bm{S_d}$ multiplied by $1$ and $-1$ are fed into several linear projection layers and GeLU functions. We expect these GeLU functions and linear projection layers can learn to truncate excessive noise realizations. To introduce the position information to the decoding process, we add the same position embedding layer as in \eqref{pos_embed} to get the output $\bm{D_{0}}\in \mathbb{R}^{l\times d}$:
\begin{equation}
\bm{D_{0}}=\text{Siam}(\bm{S_d})+P_e(\bm{p}).
\label{siam_embed}
\end{equation}


\textbf{Transformer layer:}
After the Siamese layer and positional embedding, $\bm{D_{0}}$ is passed through $L_t$ transformer layers:  
\begin{equation}
\bm{D_{l}}=MSA(\bm{D_{l-1}})+MLP(MSA(\bm{D_{l-1}})),
\end{equation}
where $\bm{D_{i}}\in \mathbb{R}^{l\times d}$ is the output of the $i$-th transformer layer at the decoder, and the $MSA$ and $MLP$ blocks share the same structure as those in \eqref{trans_layer_eqn}.

\textbf{Linear Projection:}
Given the output of the $L_t$-transformer layer $\bm{D_{L_t}}$, we apply a linear projection $\bm{W_{out}}\in \mathbb{R}^{d\times c}$, and then reshape the output into a matrix of size $\mathbb{R}^{h\times w \times 3}$ to reconstruct the input image as:
\begin{equation}
\bm{\hat{S}}=\text{reshape}(\bm{D_{L_t}}\bm{W_{out}}).
\label{last_dense_de}
\end{equation}

\subsection{Loss function}
The encoder and decoder is optimized jointly by minimizing the MSE between the input image $\bm{S}$ and its reconstruction $\bm{\hat{S}}$: 
\begin{equation}
    \mathcal{L}(\bm{\theta},\bm{\phi} )= \mathbb{E}_{\bm{\theta,\phi}}\big[ \|\bm{S}-\bm{\hat{S}}\|^2_2\big].
    \label{loss_func}
\end{equation}
We train the model to search for the optimal parameter pair $(\bm{\theta}^*,\bm{\phi}^* )$ with a minimal loss $\mathcal{L}(\bm{\theta},\bm{\phi})$ as:
\begin{equation}
(\bm{\theta}^*,\bm{\phi}^* )=\mathop{\arg\min}_{\bm{\theta},\bm{\phi}}\mathbb{E}\big[\mathcal{L}(\bm{\theta},\bm{\phi})\big],
\label{goal}
\end{equation}
where the expectation is taken over the training image and channel datasets.
\begin{figure*}[tb]
    \centering
    \subfloat[$R=1/24$]{
    \label{close_ViT_1_12} 
    \begin{minipage}[t]{0.32\linewidth}
    \centering
    \includegraphics[scale=0.33]{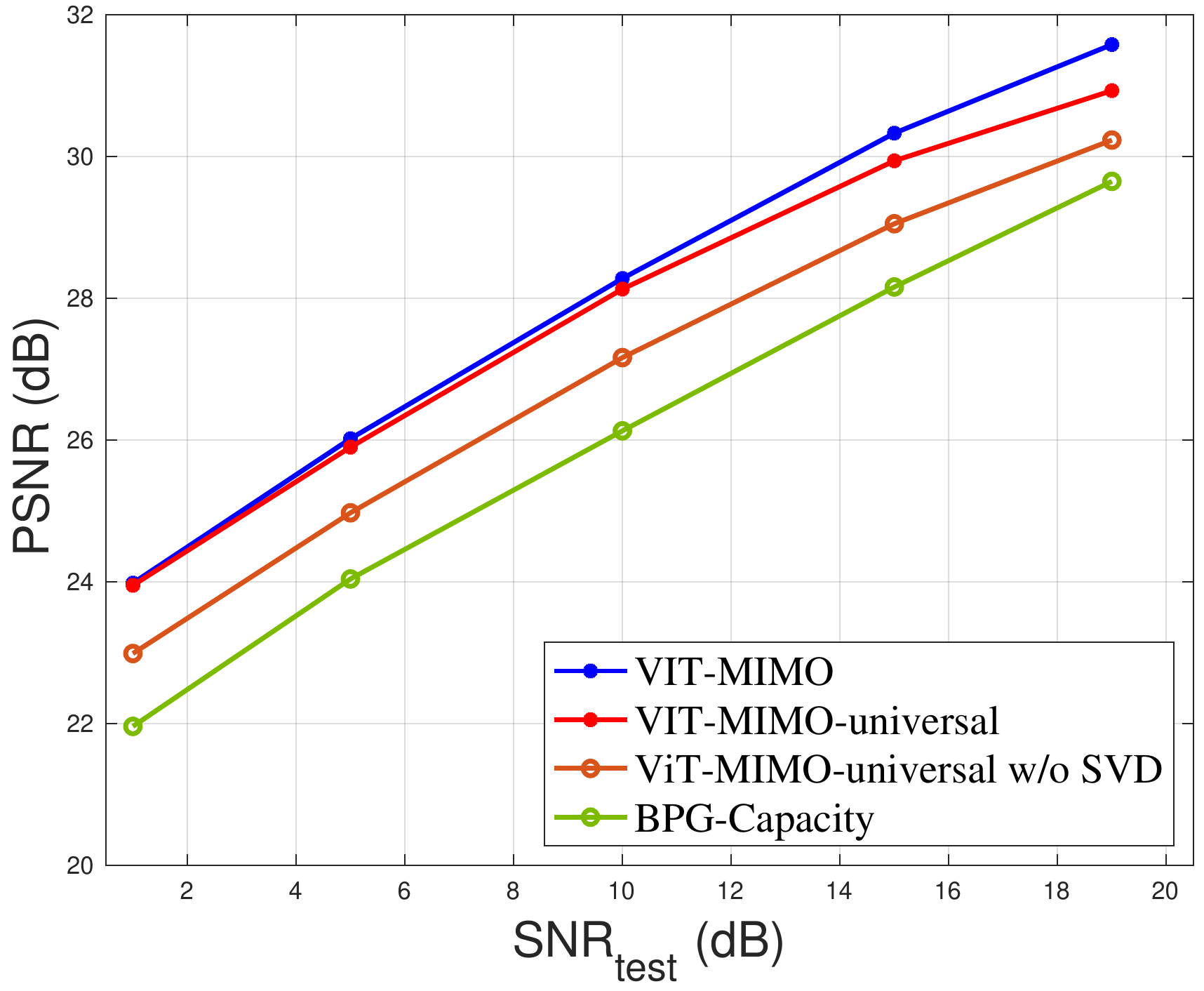}
    \end{minipage}%
    }%
    \subfloat[$R=1/12$]{
    \label{close_ViT_1_6} 
    \begin{minipage}[t]{0.32\linewidth}
    \centering
    \includegraphics[scale=0.33]{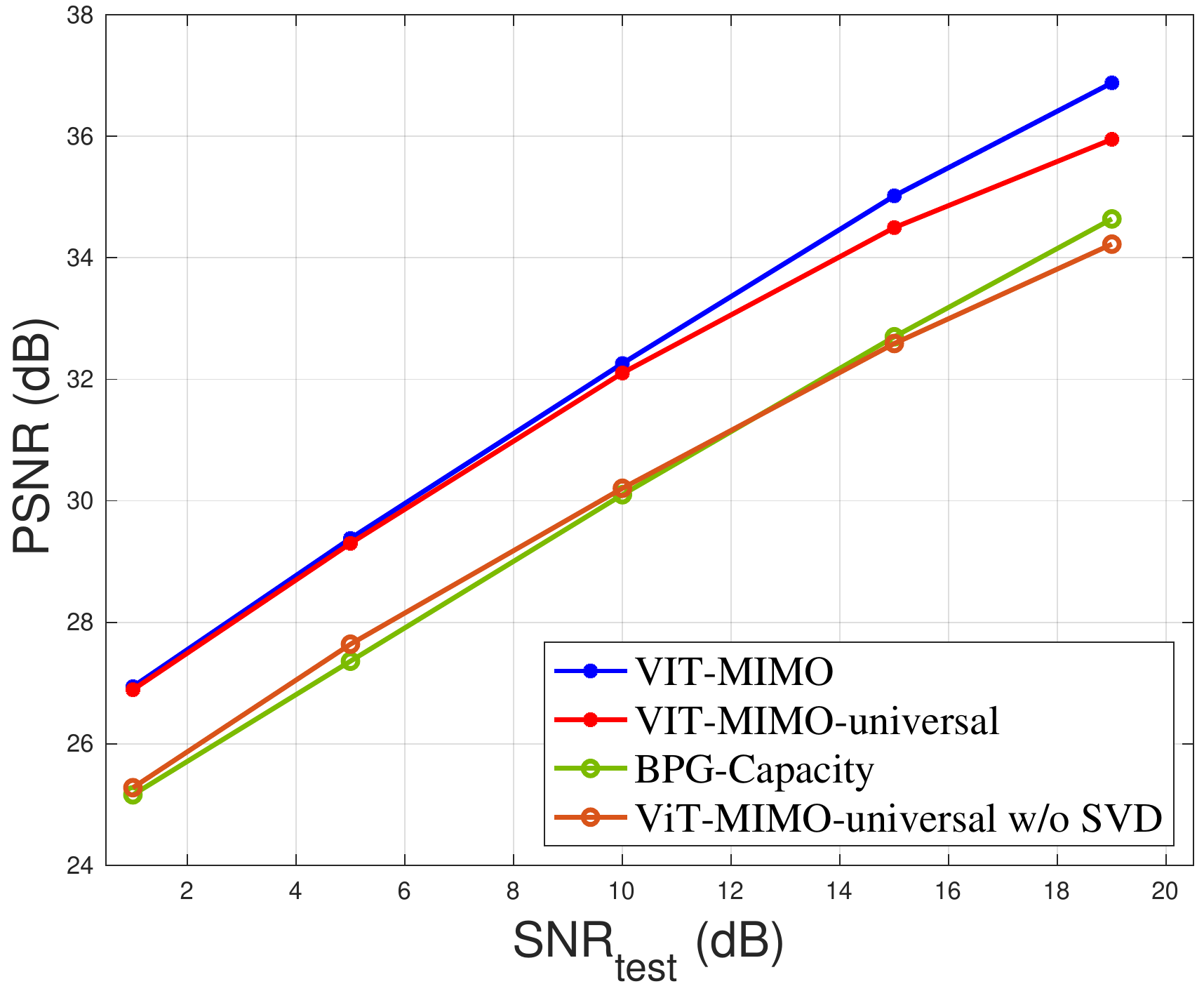}
    \end{minipage}%
    }%
    \subfloat[$R=1/6$]{
    \label{close_ViT_1_3} 
    \begin{minipage}[t]{0.32\linewidth}
    \centering
    \includegraphics[scale=0.33]{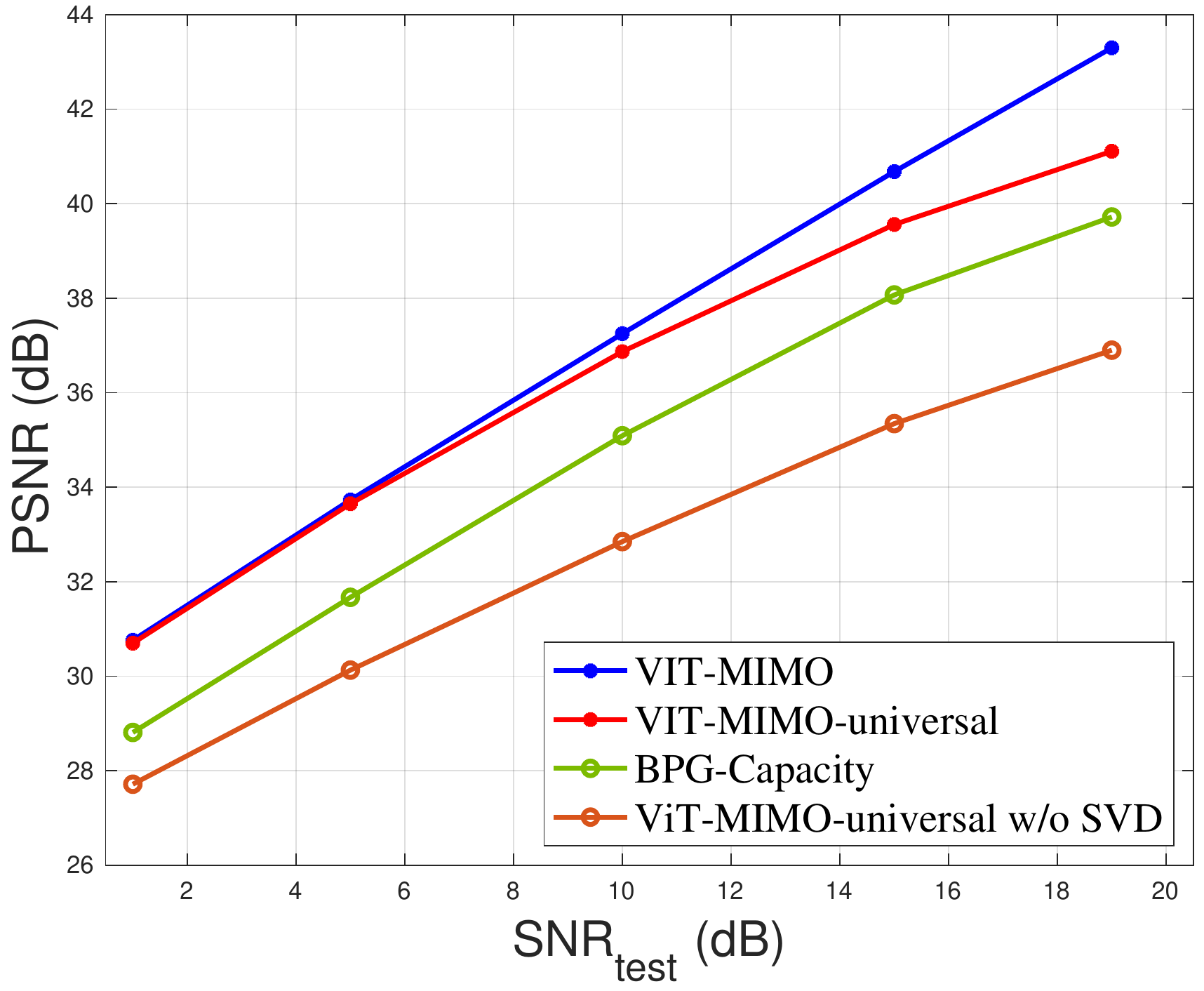}
    \end{minipage}%
    }%
    \centering
    \caption{Performance of the proposed ViT-MIMO model compared with the BPG-Capacity benchmark at different channel SNR and bandwidth ratio scenarios.}
    \label{ViT_close_JSCC} 
\end{figure*}

\section{Training and evaluation}
This section conducts a set of numerical experiments to evaluate the performance of our ViT-MIMO in various bandwidth and SNR scenarios. As a benchmark, we consider a BPG-Capacity scheme, where BPG is used for image compression, and capacity-achieving channel codes with the water-filling algorithm are assumed.
Unless stated otherwise, we consider a $2\times2$ MIMO system and the CIFAR10 dataset\cite{krizhevsky2009learning}, which has $50000$ color images of size $32\times 32\times 3$ in the training dataset and $10000$ images in the test dataset.

Our model is implemented in Pytorch with two GTX 3090Ti GPUs. We use a learning rate of $5e^{-5}$ and a batch size of $128$ with the Adam optimizer. Considering the model complexity and experiment performance, we set $p=8$, $l=64$, $c=48$ for image vectorization, and $L_t=8$, $d=256$, and $N_s=8$ for each transformer layer of the ViT. Each element of $\bm{H}$ follows $H[i,j]\sim \mathcal{CN}(0,1)$. The average received SNR $\mu$ is defined as:
\begin{equation}
    \mu\triangleq 10\log_{10} \frac{\mathbb{E}_{\bm{H,X}}[\bm{\|HX\|}^2_F]}{\mathbb{E}_{\bm{W}}[\bm{\|W\|}^2_F]}~(\text{dB})=10\log_{10} \frac{M}{\sigma_w^2}.
\end{equation}

\subsection{General performance}
We first evaluate the ViT-MIMO model under the setup where the training SNR, denoted by $\textit{SNR}_{\textit{train}}$ matches the test SNR, $\textit{SNR}_{\textit{test}}$. In particular, we set $\textit{SNR}_{\textit{train}}=\textit{SNR}_{\textit{test}}\in \{1,5,10,15,19\}$dB and the bandwidth ratio $R\in\{1/24,1/12,1/6\}$.

The comparison between ViT-MIMO and the BPG-Capacity benchmark is shown in Fig. \ref{ViT_close_JSCC}.
As can be seen, ViT-MIMO outperforms the separation-base benchmark in all SNR and bandwidth-ratio scenarios. Specifically, from Fig. \ref{close_ViT_1_12} and \ref{close_ViT_1_6}, we can see that ViT-MIMO significantly outperforms the benchmark by at least $1.98$dB and $1.78$dB when $R=1/24$ and $R=1/12$, respectively. When $R=1/6$, improvements of up to $3.5$dB can be observed at all test SNRs. These significant improvements demonstrate the superiority of our ViT-MIMO scheme, and its capacity in extracting and mapping image features to the available channels in an adaptive fashion.



\begin{figure}[t]
     \centering
    \includegraphics[scale=0.38]{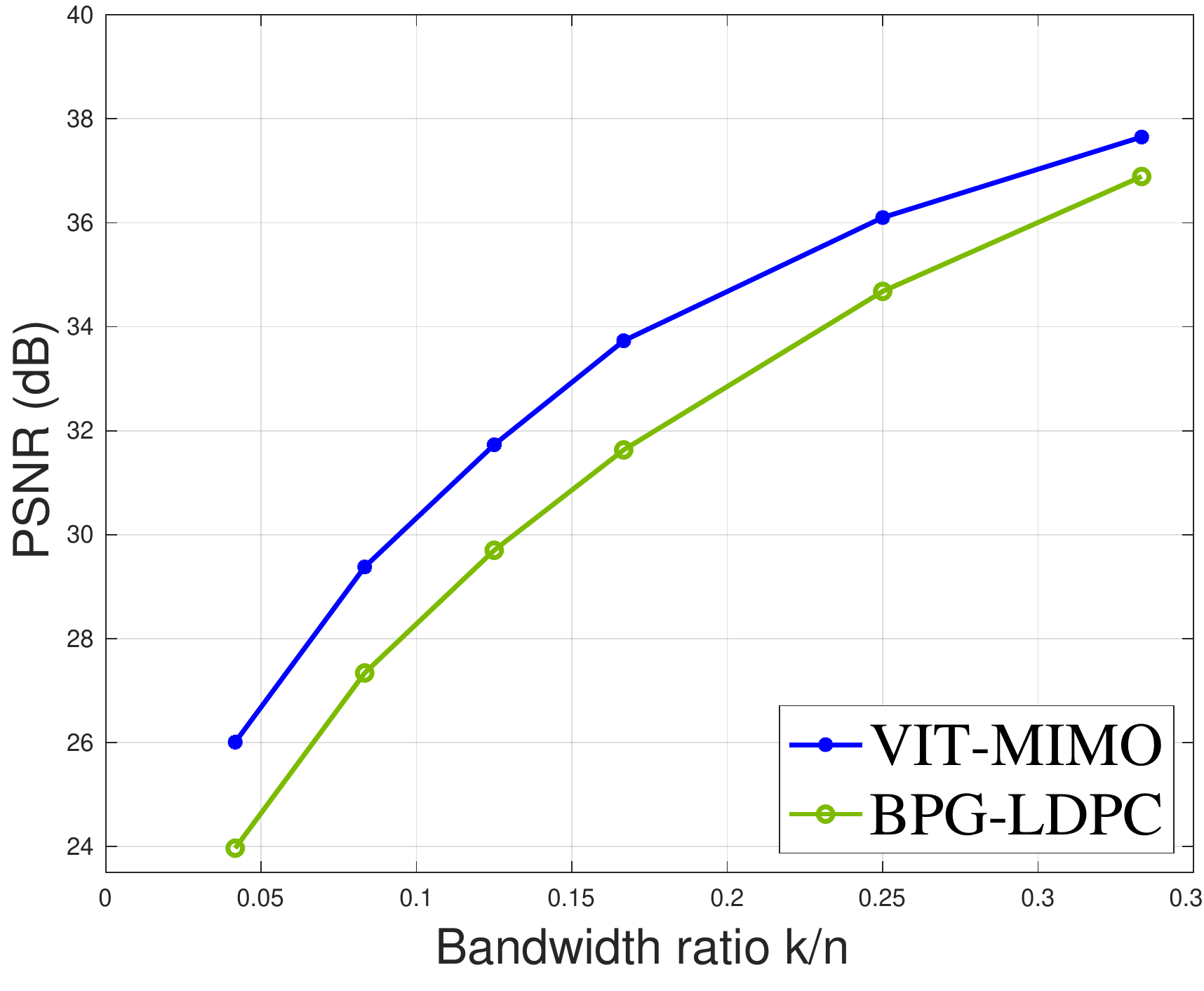}
    \caption{Performance of different schemes over different bandwidth ratios.}
    \label{close_bd_5}
 \end{figure}
 
\subsection{Different bandwidth ratios}
To further evaluate the impact of the bandwidth ratio $R$ on the system performance, we compare the ViT-MIMO and BPG-Capacity schemes over a wide range of bandwidth ratios in Fig. \ref{close_bd_5}, where the test SNR is set to $5$dB. We observe that ViT-MIMO can outperform the BPG-Capacity benchmark at all bandwidth ratios, with a gain of up to $2.08$dB in PSNR. The gap between the two is more significant for low $R$ values. We would like to emphasize that the BPG-Capacity performance shown in these figures is not achievable by practical separate source and channel coding schemes, and the gap can be significant especially in the shorter block lengths, i.e., low $R$. Thus, the comparison here illustrates the clear superiority of ViT-MIMO under various bandwidth ratios compared to separation-based alternatives. 

\subsection{SNR-adaptability}
We also consider a random SNR training strategy to evaluate the channel adaptability of the ViT-MIMO scheme. We train the ViT-MIMO model with random SNR values uniformly sampled from $[0,22]$ dB and test the well-trained model at different test SNRs, denoted by ViT-MIMO-universal. The comparisons in Fig. \ref{ViT_close_JSCC} show that there is a slight performance degradation compared to training a separate ViT-MIMO encoder-decoder pair for each channel SNR (up to $0.65$dB, $0.83$dB and $2.1$dB for $R=1/24$, $R=1/12$ and $R=1/6$); however, the ViT-MIMO-universal brings significant advantage in terms of training complexity and storage requirements.

We can observe more performance loss in high SNR regimes. However, our ViT-MIMO-universal still significantly outperforms the BPG-Capacity benchmark. We conclude that the ViT-MIMO-universal scheme can adapt to SNR-variations with a slight loss in PSNR, especially in the high-SNR regime.

\subsection{SVD ablation study}
To evaluate the effectiveness of the SVD strategy in our scheme, we train the models of ViT-MIMO-universal without SVD decomposition-based precoding in different bandwidth ratios, denoted by ViT-MIMO-universal w/o SVD. Specifically, we feed the channel heatmap $\bm{M}$ into both transceivers, and 
the performance is shown in Fig. \ref{ViT_close_JSCC}. Compared with the ViT-MIMO-universal, we can observe that the performance of ViT-MIMO-universal w/o SVD is sub-optimal with the performance loss exceeding $0.97$dB, $1.6$dB and $2.9$dB for $R=1/24$, $R=1/12$ and $R=1/6$, respectively. We can conclude that the SVD-based model-driven strategy can significantly simplify the training process and improve transmission performance, which illustrates the importance of exploiting domain knowledge in the design of data-driven communication technologies.

\section{Conclusion}
This paper presents the first DeepJSCC-enabled MIMO communication system for image transmission, where a vision transformer is designed to explore the contextual semantic features and the channel conditions with a self-attention mechanism. Numerical experiments show that our ViT-MIMO method can significantly improve the transmission quality under various SNRs and bandwidth scenarios. In addition, the proposed ViT-MIMO is a unified and compact design that can adapt to channel variations without retraining.

Moving forward, we will consider extending our ViT-MIMO scheme to the scenario in which the CSI is available only to the receiver. We will also explore new deep learning-aided MIMO equalization methods and space-time coding to improve the performance. 
\bibliographystyle{IEEEtran}
\bibliography{ref}
\end{document}